# Enhancement of Spin Coherence using *Q*-factor engineering in Semiconductor Microdisk Lasers

S. Ghosh[1], W. H. Wang[2], F. M. Mendoza[1], R. C. Myers[1], X. Li[2], N. Samarth[2], A. C. Gossard[1], and D. D. Awschalom[1].

[1]*Center for Spintronics and Quantum Computation, University of California, Santa Barbara, CA 93106, USA*

[2]*Materials Research Institute, Penn State University, University Park, PA 16802, USA*

**Semiconductor microcavities offer unique means of controlling light-matter interactions, which have led to the development of a wide range of applications in optical communications[1] and inspired proposals for quantum information processing and computational schemes[2,3]. Studies of spin dynamics in microcavities – a new and promising research field – have revealed novel effects such as polarization beats, stimulated spin scattering, and giant Faraday rotation[4-7]. Here, we study the electron spin dynamics in optically-pumped GaAs microdisk lasers with quantum wells (QWs) and interface-fluctuation quantum dots[8] (QDs) in the active region. In particular, we address the question of how the electron spin dynamics are modified by the stimulated emission in the disks, and observe an enhancement of the spin lifetime when the optical excitation is in resonance with a high quality (*Q* ~ 5000) lasing mode. This resonant enhancement, contrary to what is expected from the Purcell effect[9] observed in the cavities, is then manipulated by altering the cavity design and dimensions.**

Semiconductor microdisk lasers, based on whispering gallery (WG) optical resonances[10], have been the subject of considerable interest, given their fast response time, scalability and in-plane emission[11]. These features make them attractive components for on-chip integration in optoelectronic devices. The fabrication of microdisks embedded with quantum dots[12] has led to proposals of quantum computational schemes based on the mediation of interactions between distant QD spins coherently coupled through a single microdisk mode[2]. In this letter, we study the coupling between the microdisk emission and the localized spins in its active region, as a first step towards the realization of using light-matter information exchange for quantum information processing schemes in a solid-state system.

Our samples are fabricated from GaAs/AlGaAs heterostructures grown by molecular beam epitaxy. Figure 1a (inset) is a scanning electron microscope (SEM) image of a representative microdisk with a diameter ~ 4 μm. The optically active region contains five 4.2 nm thick GaAs QWs, separated by 10 nm thick $Al_{0.31}Ga_{0.69}As$ barriers. The total thickness of the disk is ~ 110 nm, ensuring that only the lowest order cavity modes are supported[10]. At each QW/barrier interface, growth interruptions of 2 minutes are introduced to induce monolayer fluctuations, leading to the formation of QDs. Control samples, consisting of disks of identical dimensions, are fabricated from heterostructures with the same design but with only 5 seconds of growth interruptions. Low temperature micro-photoluminescence measurements at low excitation power ($P <$ 10 μW) show clear, resolution-limited spectral features (FWHM ~ 200 μeV) only in the former samples, confirming the presence of QDs in the active region[13]. At higher excitation powers both samples support WG resonant modes, but the $Q$-factor is consistently higher in the QD samples ($Q$ ~ 5500) than in the control samples ($Q$ ~ 2000).

Spectroscopic and dynamic details of the stimulated emission at temperature $T$ = 5.5 K from a microdisk with QDs are shown in Fig.1. A tunable, pulsed Ti:Sapphire laser, producing ~150 fs wide pulses at a repetition rate of 76 MHz, is tuned to 740 nm to optically pump the disks for photoluminescence (PL) measurements. The samples are mounted in an optical cryostat with a variable temperature insert and a 7 T superconducting magnet. The incident laser is focused to a spot ~ 30 μm in diameter at an angle of 45 degrees to the sample normal. The PL is collected perpendicular to the excitation direction in reflection geometry. The inter-disk distance is 50 μm, allowing us to address a single structure within the laser spot. At low incident power ($P$ < 200 W/cm$^2$) the emission is mostly spontaneous, with the beginnings of cavity coupling exhibited through the emergence of a few modes. With increasing power, one of these modes starts to dominate, and the spectrum shows resonant emission at 770 nm with a transparency $Q$ = 5500. Figure 1a shows the PL emission of the mode at $P$ =1.3 kW/cm$^2$ and Fig. 1b, the contrast between the power dependence of the time-integrated intensities of the lasing mode and the spontaneous emission. A fit to the logarithmic plot of the mode output intensity as a function of input power[14] indicates stimulated emission with a threshold of 450 W/cm$^2$. Fig. 1c shows time-resolved PL traces of the spectrally-resolved lasing emission at different pump powers, taken with a streak camera (time resolution ~ 2 ps). Above the threshold power, the delay time, defined as the time interval between the arrival of the laser pulse (marked $t$ = 0) and the maximum of the cavity emission, starts decreasing with increasing pump power, as shown in Fig. 1b. This is another indication of the onset of stimulated emission[15]. The decrease in the recombination time is a result of the Purcell effect[16].

Time-resolved Kerr rotation (TRKR), an optical pump-probe spectroscopic technique[17,18], is used to probe the electron spin dynamics in the cavities (in the Voigt geometry, detailed in Fig. S1a). A circularly-polarized pump pulse, incident on the sample surface, injects spin-polarized electrons, with the circular polarization being modulated with a photoelastic modulator at 50 kHz for lock-in detection. The Kerr rotation angle of a linearly-polarized probe pulse (tuned to the same wavelength as the pump) applied after a time delay $\Delta t$ measures the projection of the electron spin magnetization as it precesses about the applied magnetic field. We extract a spin lifetime, $T_2$*, from the exponentially decreasing envelope of the spin precession curve and the Larmor precession frequency, $\omega_L$, from the transverse oscillations (Fig. S1b).

Figure 2a is a map of $T_2$*, at $T$ = 5.5 K with an applied magnetic field $B$ = 4 T, as we vary both the pump wavelength ($\lambda$) and power ($P$). We immediately notice an anomaly near the lasing wavelength at $\lambda$ = 770 nm. $T_2$* first increases to 456 ps when $P$ ~ 0.5 kW/cm$^2$, and then decreases to 158 ps as $P$ is increased to 1.5 kW/cm$^2$. Such drastic variations in $T_2$* are not observed at any other wavelengths, as shown further by the line-cuts taken at fixed $\lambda$ (770 nm and 775 nm) and plotted in Fig. 2b. For comparison, we also plot $T_2$* as a function of $P$ for the unprocessed heterostructure (with QDs but no cavity) and for a resonant mode in a cavity in the control sample (no QDs). Neither of these shows a comparable variation of $T_2$* with $P$ as seen in the QD sample. While this is expected in the unprocessed heterostructure, the lack of resonant enhancement in the control sample may be attributed to the lower quality of the mode.

A careful look at Fig. 2a reveals that the value of $P$ where the increase in $T_2$* occurs in the QD sample is very close to the threshold power for lasing in the disk. The lasing threshold is characterized by changes in various parameters, such as an increase in the mode output intensity, a decrease in the delay time, and a change of the mode $Q$-factor as the system crosses over from spontaneous to stimulated emission. Plotting $Q$ of the resonant mode (obtained from PL) as a function of $P$ in Fig. 2c shows an increase at the threshold and a subsequent decrease at higher powers. The qualitative variation of $Q$ with $P$ traces the evolution of the spin lifetime very well, revealing the possibility that $T_2$* can be enhanced by engineering cavities with larger $Q$. Figure 2d shows the Larmor frequency $\omega_L$ in the QD sample at $B$ = 4 T with $\lambda$ = 770 nm and 775 nm (on and off resonance, respectively) as $P$ is varied. It also increases at the resonant wavelength, over the same range in power as $T_2$*.

In addition to varying $P$, we may also vary the temperature to alter the lasing characteristics of the microdisk. In Fig. 3a we map out $T_2$* as a function of $P$ at different temperatures with the pump fixed at 770 nm. The modulations in $T_2$* with $P$ decrease with temperature, disappearing by $T$ = 35 K, shown by the line-cuts at three different temperatures (Fig. 3b). In Fig. 3c we follow the change in $T_2$* with temperature at $P$ = 0.5 kW/cm$^2$ (black) and once again, it agrees qualitatively with the variation of the mode $Q$-factor (red) with temperature; both decrease exponentially. Not only is the mode quality degraded, but the emission at 770 nm is no longer dominated by stimulated emission, evident in the sharp increase in the delay time around 35 K (Fig. 3d).

Due to the spectral overlap of the emission from the QWs and the QDs, it is not possible to unequivocally identify either as the sole source of the lasing emission in the disks. However, the samples with QDs have (a) consistently higher $Q$-factors, implying lower dissipation, and (b) shorter delay times, resulting from less carrier diffusion to the edges. This suggests the involvement of the QDs, which provide confinement of the carriers and hence prevent non-radiative recombination at the sidewalls, at least under conditions of “moderate” excitation and low temperature. As the pump power (temperature) increases, QD states saturate (depopulate) and recombination is favoured from QW states. The carriers not only diffuse to the sidewalls and recombine non-radiatively, but also begin to couple extensively to radial modes, which is borne out by the PL data[13], where we see “mode hopping” – the lasing emission cascading to longer wavelengths by coupling to second and third order radial modes (which are far more dissipative, and consequently, have lower $Q$) as $T$ and $P$ increase.

In an attempt to reduce the losses at higher excitation powers and temperature, we designed and fabricated disks with a smaller diameter (~ 1.5 µm). Fig. 4a is an SEM image of such a disk. The chief advantage of these smaller disks is that the spectral width of the gain supports only one mode (the mode spacing of the WG modes is inversely proportional to the radius of the disk) and should, in principle, lead to reduced losses. PL emission from the disk consists of a single lasing mode at 776 nm, with a *Q*-factor that increases at low powers. However, in sharp contrast to larger disks, the *Q*-factor remains nearly constant at ~ 4500 over a large power range (Fig. 4c, red squares), showing no indication of mode quality degradation or emergence of higher order radial modes. The spin lifetime plotted in Fig. 4b, exhibits resonant enhancement at the lasing mode, and also does not decrease with increasing power, unlike in the larger disks. It persists until the pump power is quite high (2.2 kW/cm$^2$). The line-cuts in Fig. 4c show $T_2$* on- and off-resonance, and in the unprocessed heterostructure. $T_2$* in the cavity once again follows the variation in mode *Q*-factor. The Larmor precession frequency $\omega_L$ for two different pump wavelengths is shown in Fig. 4d, and it is found to be larger on-resonance than off.

In summary, we have demonstrated the possibility of increasing spin lifetime at selective wavelengths in a cavity, by coupling to high-*Q* modes. This resonant enhancement would be of great interest in microdisks with embedded self-assembled quantum dots, for which very high *Q* values have been reported[19]. At this point, we do not have an explanation for the resonant increase in the spin lifetime or the Larmor precession frequency, although it is a clear indication of the existence of some form of coupling between the cavity emission and the carrier dynamics. However, our results show that we can engineer control over the cavity-spin coupling via microcavity designs that robustly enhance the spin lifetime by reducing mode degradation. Recent results have shown the possibility of creating entanglement between resonant states of optically coupled disks[20], and of efficient read-out by fibre coupling to disks[21]. These advancements taken together make these systems very promising candidates for quantum information processing.

‘**Supplementary Information** accompanies the paper on **www.nature.com/nmat**.’

The authors would like to acknowledge support from DARPA/QUIST and NSF, and thank E. L. Hu, R. J. Epstein and F. Meier for illuminating discussions.

Correspondence and requests for materials should be addressed to D. D. A. at awsch@physics.ucsb.edu.

**Figure 1 Static and dynamical characteristics of microdisks. a**, Photoluminescent (PL) spectral emission from a disk (diameter ~ 4 µm) excited by a pulsed laser at 740 nm, exhibiting a resonant mode at 770 nm with $Q$ ~ 4800. (inset) SEM image of a disk. **b**, Time-integrated intensity of the resonant emission (red) and the spontaneous emission (black triangles) as a function of pump power $P$. Fit to the mode intensity (red line) yields the lasing threshold. **c**, Time-resolved emission at 770 nm with varying pump power. The delay time decreases with increasing input power, starting at the threshold, as shown in **b** (black circles).

**Figure S1 Time-resolved Kerr rotation. a**, Set-up for time-resolved Kerr rotation. The circularly-polarized pump and linearly-polarized probe are incident along $z$, perpendicular to the magnetic field $B$ along $x$. The time delay $\Delta t$ between the pump and the probe is varied to trace out the variation in the Kerr rotation angle $\theta_K$. The sample geometry is the same as for the PL measurements. **b**, Time-resolved spin dynamics measured by Kerr rotation spectroscopy. The pump and probe are resonantly tuned to the cavity mode at 770 nm. The spin lifetime ($T_2$*) varies non-monotonically with incident pump power ($P$).

**Figure 2 Resonantly enhanced spin coherence. a**, Map of $T_2$* as a function of pump wavelength ($\lambda$) and power ($P$). Spin coherence is enhanced at the lasing mode at low pump powers, and decreases as the power increases. **b**, $T_2$* as a function of pump power from line-cuts across a, illustrating the resonant coupling at 770 nm (black). A similar line cut at 775 nm (red), off-resonance, shows no such enhancement. $T_2$* in the unprocessed heterostructure (open circles), as well as in a control sample (blue), both shown here at 770 nm, remain unchanged with $P$. **c**, Variations in $T_2$* and the cavity $Q$-factor at 770 nm with pump power. **d**, The Larmor precession frequency, $\omega_L$, in the cavity, obtained from fits to spin precession data, as a function of $P$. For all the Kerr rotation measurements, the probe power is ~ 15 $W/cm^2$.

**Figure 3 Temperature dependence. a**, $T_2$* as a function of temperature ($T$) and pump power ($P$) at the cavity resonance (770 nm). **b**, Line cuts across **a**, at $T$ = 5, 20 and 30 K, showing the resonant enhancement of $T_2$* diminishing with increasing *T*. **c**, $T_2$* (black, $\lambda$ = 770 nm, *P* = 0.5 kW/cm$^2$) decreases with *T.* Variation of the mode *Q*-factor (red), and of the delay time of disk emission (**d**), point to the decrease being related to the reduction of stimulated emission with increasing temperature. Lines in **c** are exponential fits.

**Figure 4 Robust spin-coherence enhancement in smaller cavity. a**, SEM image of microdisk with diameter ~ 1.5 µm. This reduced size causes the resonance wavelength to shift to 776 nm. **b**, $T_2$* as a function of pump wavelength ($\lambda$) and power (*P*). The resonant increase in $T_2$* at 776 nm is robust to higher values of *P*. **c**, Line cuts across b for 776 nm (black) and 771 nm (blue) along with the power dependence of $T_2$* in the unprocessed heterostructure (red circles) and of *Q* (red squares). **d**, $\omega_L$ at 776 and 771 nm, showing a similar trend with *P* as $T_2$*.

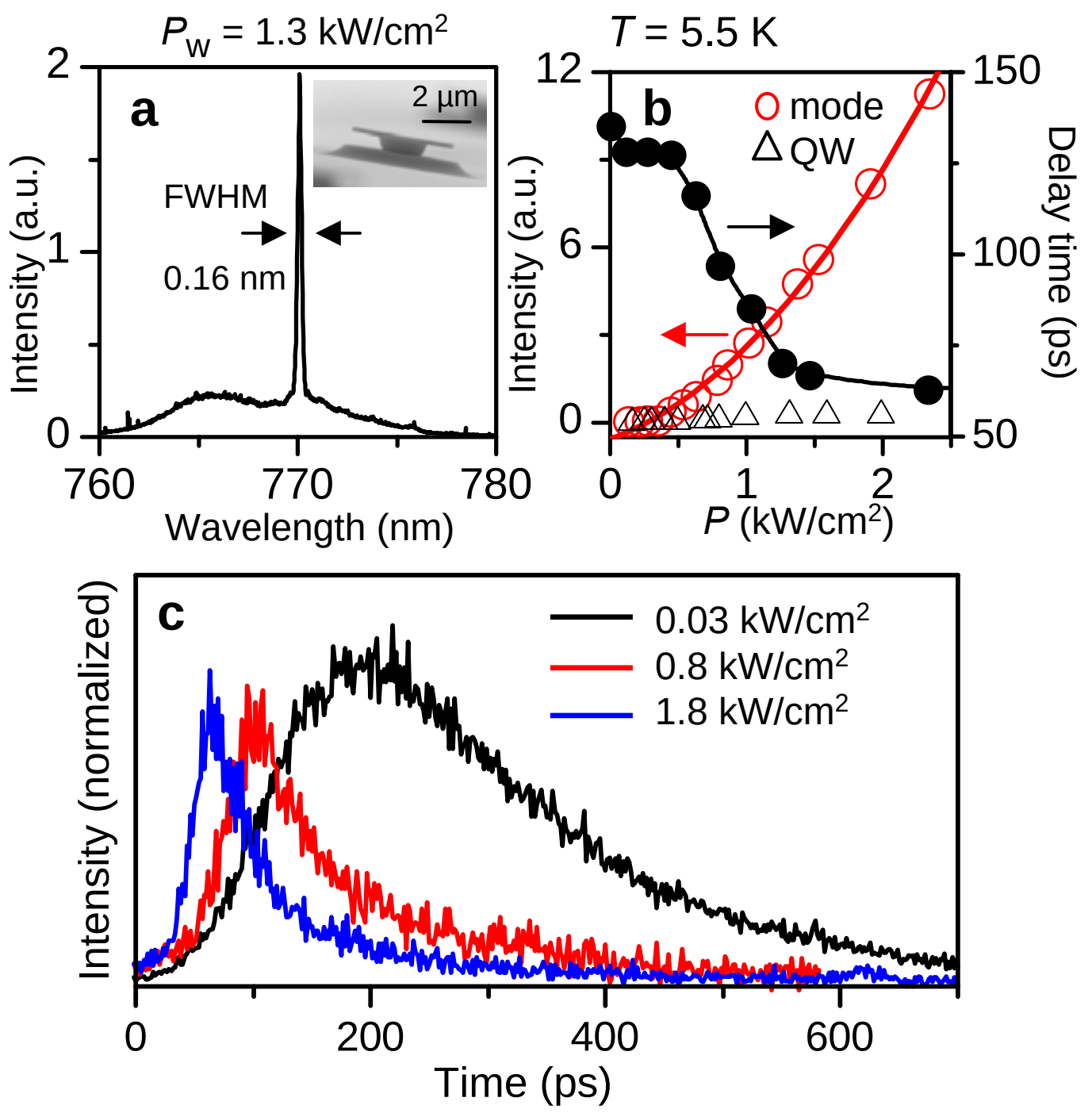

Figure 1

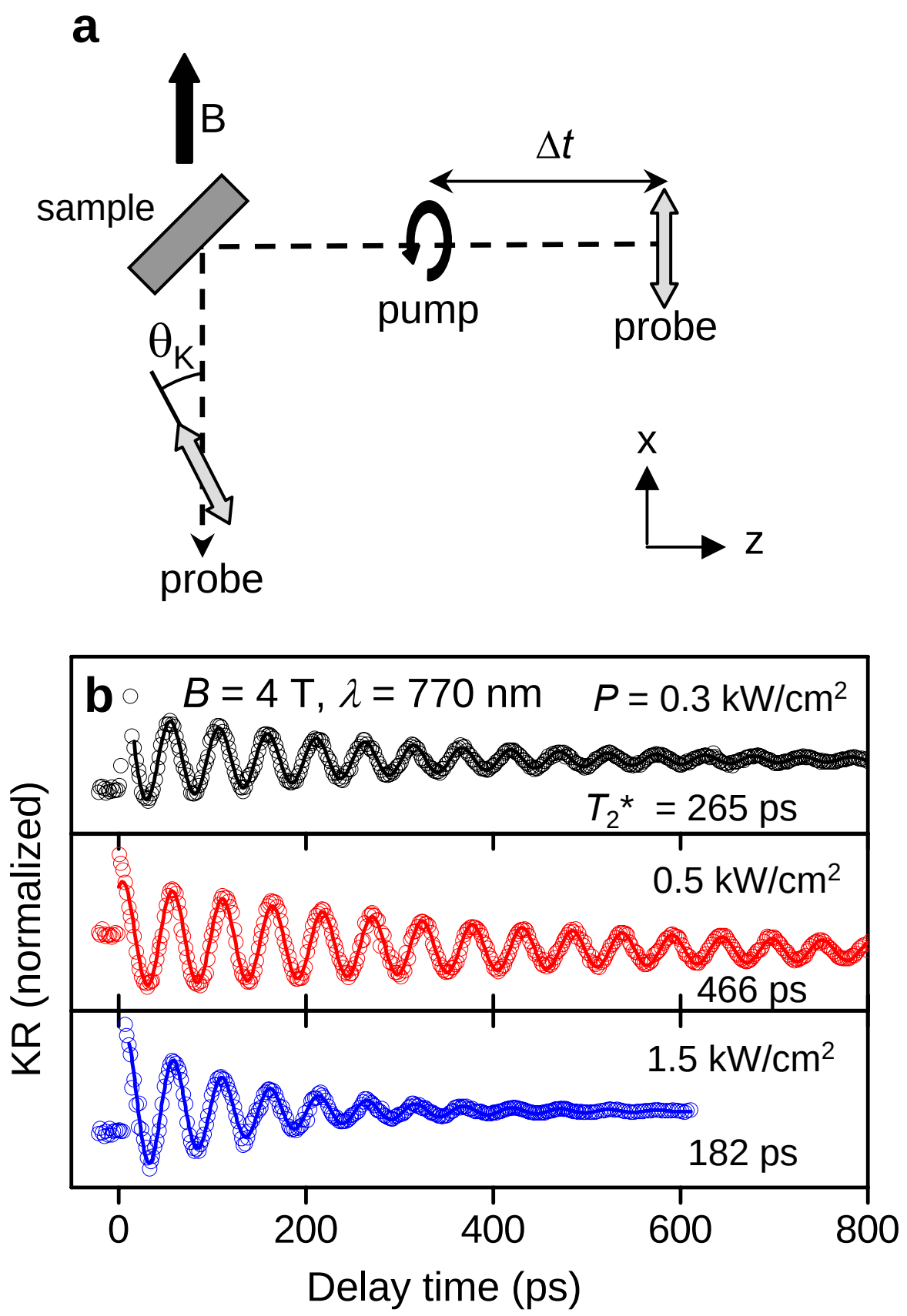
a
B
sample
Δt
pump
probe
θK
probe
x
z
b
B = 4 T, λ = 770 nm
P = 0.3 kW/cm2
T2* = 265 ps
0.5 kW/cm2
466 ps
1.5 kW/cm2
182 ps
KR (normalized)
0
200
400
600
800
Delay time (ps)

Figure S1

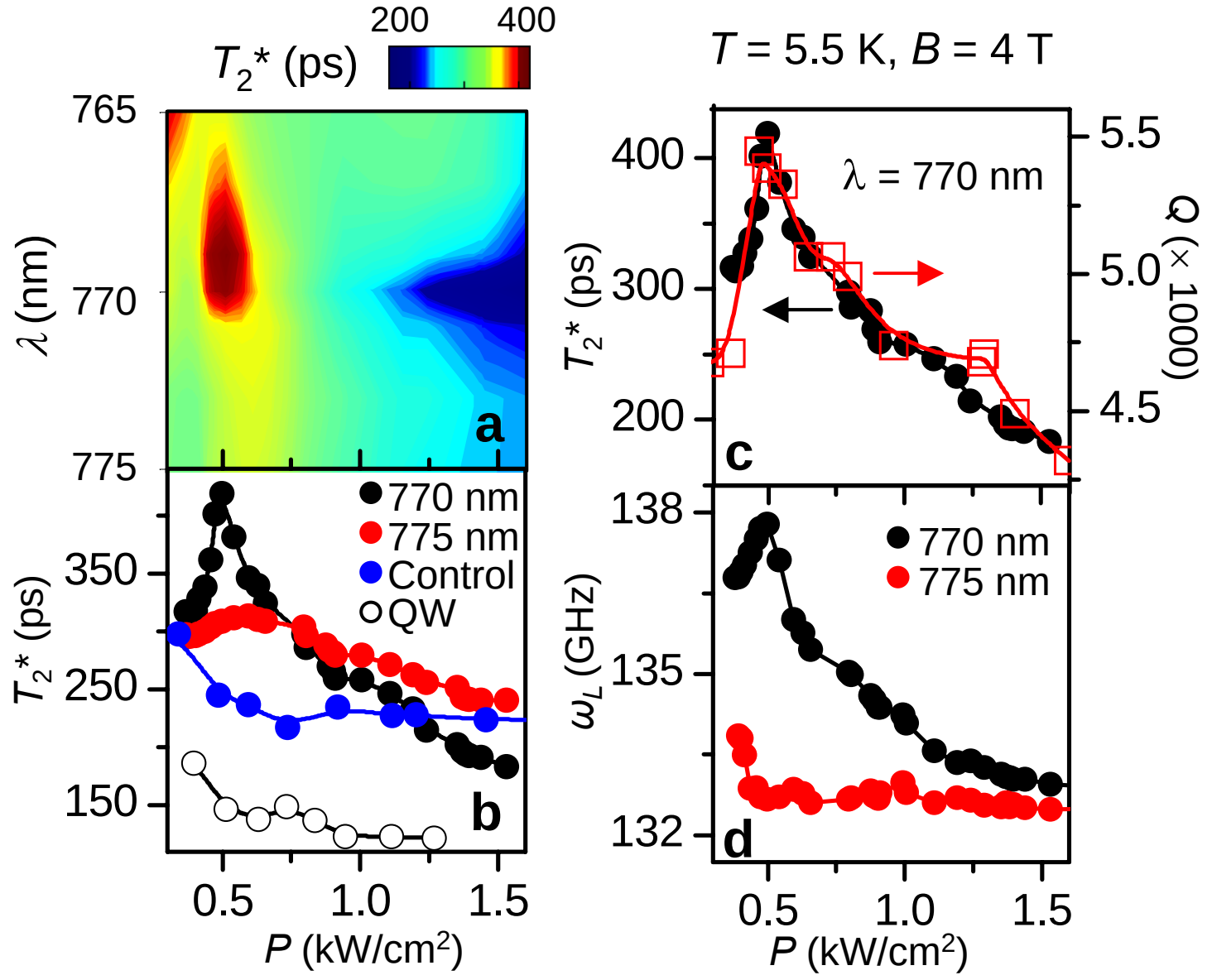

T_2* (ps)
200
400
765
770
775
λ (nm)
a
770 nm
775 nm
Control
QW
350
250
150
T_2* (ps)
b
0.5
1.0
1.5
P (kW/cm²)
T = 5.5 K, B = 4 T
λ = 770 nm
400
300
200
T_2* (ps)
5.5
5.0
4.5
Q (× 1000)
c
138
135
132
ω_L (GHz)
770 nm
775 nm
d
0.5
1.0
1.5
P (kW/cm²)

Figure 2

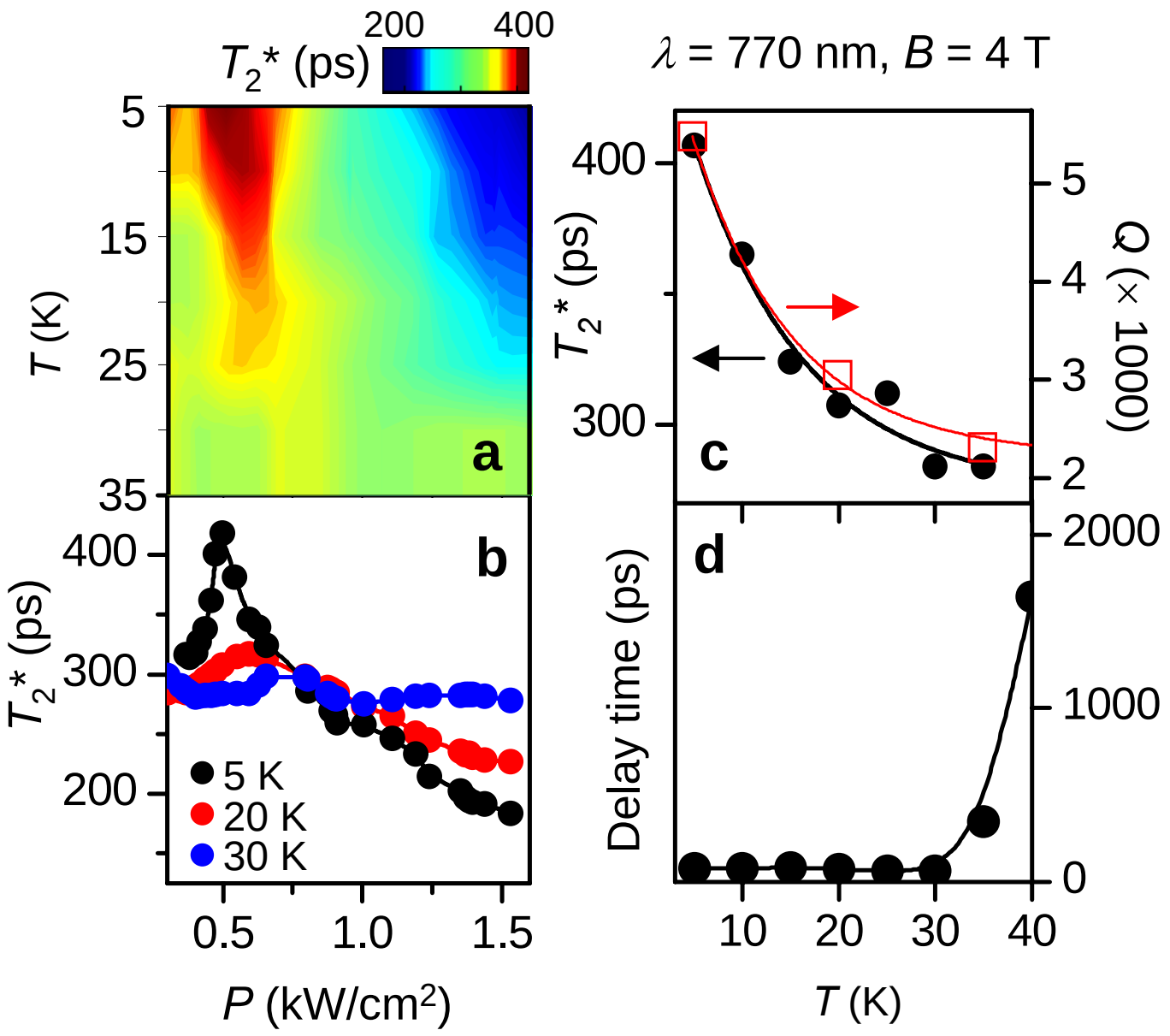

T2* (ps)
200
400
5
15
25
35
T (K)
a
b
400
300
200
T2* (ps)
5 K
20 K
30 K
0.5
1.0
1.5
P (kW/cm2)
λ = 770 nm, B = 4 T
400
300
T2* (ps)
c
5
4
3
2
Q (× 1000)
d
Delay time (ps)
2000
1000
0
10
20
30
40
T (K)

Figure 3

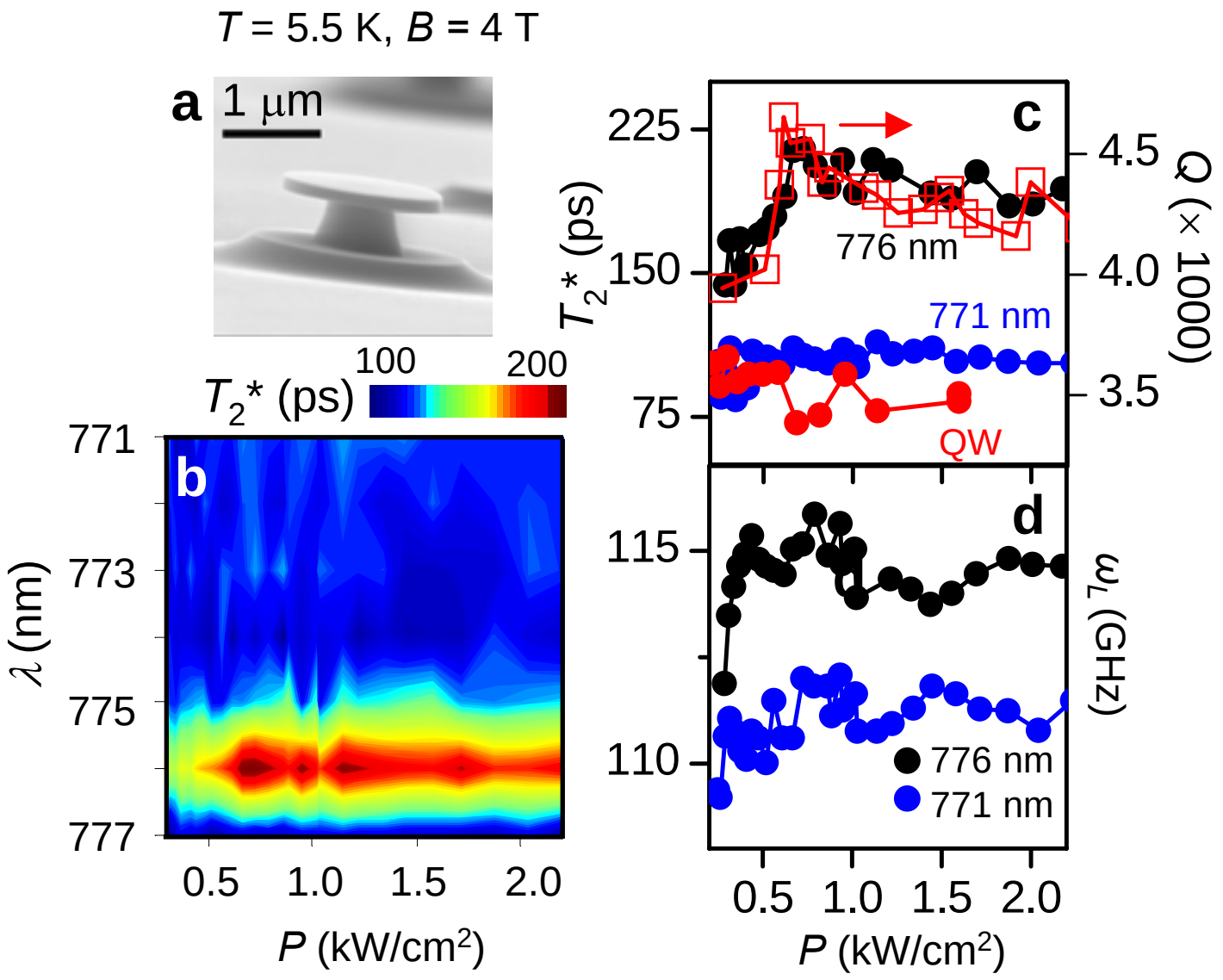

T = 5.5 K, B = 4 T
a
1 µm
b
T2* (ps)
100
200
771
773
775
777
λ (nm)
0.5
1.0
1.5
2.0
P (kW/cm2)
c
225
150
75
T2* (ps)
4.5
4.0
3.5
Q (× 1000)
776 nm
771 nm
QW
d
115
110
ωL (GHz)
776 nm
771 nm
0.5
1.0
1.5
2.0
P (kW/cm2)

Figure 4